\documentclass[twocolumn,amssymb,fleqn] {revtex4}
\usepackage{epsfig,amssymb,amsmath,subfigure,hyperref  }  
\usepackage{siunitx}
\usepackage{mathrsfs}
\usepackage{array}
\usepackage{lipsum}
\usepackage{color}
\usepackage{multirow}
\usepackage{tabularx}
\usepackage{caption}
\usepackage{bm}
\usepackage{graphicx}
\usepackage{times}

\begin{document}
	\captionsetup[figure]{labelfont={bf},name={FIG. },labelsep=period,justification=raggedright}
	
	\title{Discrete spacetime crystal: Intertwined spacetime symmetry breaking in a driven-dissipative spin system} 
	\author{Xingdong Luo}
	\email{luo-xingdong@sjtu.edu.cn}
	\affiliation{Wilczek Quantum Center and Key Laboratory of Artificial Structures and Quantum Control, School of Physics and Astronomy, Shanghai Jiao Tong University, Shanghai 200240, China} 
	\begin{abstract}

Non-equilibrium driving systems provide fertile ground for exploring intriguing spontaneous symmetry breaking phenomena. In this study, we report on the intertwined discrete spacetime translational symmetry breaking in a driven-dissipative spin system without pure spatial translational symmetry, resulting in the emergence of an  out-of-equilibrium nonmagnetic phase termed the "discrete spacetime crystal."	In contrast to the previously widely studied discrete time crystal, the spins establish a distinctive long-range crystalline order not only in time, but also in the intertwined spacetime direction. We further demonstrate the exponentially-long lifetime of the spatial-temporal order against generic weak environmental noise fluctuations. Finally, possible experimental realizations of our model are discussed.  Our findings shed light on the existence of exotic phases of matter characterized by intertwined space-time symmetry breaking, paving the way for future investigations into the properties and potential applications of these exotic states. 
	\end{abstract}
	\maketitle
	\section{Introduction}
Spontaneous symmetry breaking (SSB) plays a fundamental role in the classification of phases of matter. Crystals emerge when continuous spatial translational symmetry is spontaneously broken, while breaking rotational or gauge symmetry can give rise to ferromagnets and superfluids. Driving a system out of equilibrium can further extend the possibility of spontaneous symmetry breaking into the time domain, significantly enriching the phases of matter. In periodical driven systems, the discrete time translational symmetry (DTTS) of the driving Hamiltonian can also be spontaneously broken, leading to the formation of discrete time crystals. 

Discrete time crystal phase has attracted considerable attention and has been intensively studied in various physical systems over the years\cite{choi2017observation,zhang2017observation,frey2022realization,autti2018observation,kessler2021observation,mi2022time,trager2021real,zhu2019dicke,gong2018discrete,yang2021dynamical,hu2023solvable,surace2019floquet,huang2018clean,russomanno2017floquet,munoz2022floquet,yao2017discrete,collado2021emergent,von2016absolute}. Previous research of discrete time crystals has primarily focused on the driven systems where the temporal periodicity is treated separately from the spatial one, whereby the space-time unit cell is simply constructed by the direct product of discrete spatial and temporal translational symmetries.  In fact, more generic spacetime symmetry structures can be achieved by proper design of the driving protocol. In a pioneering work by Ref.\cite{xu2018space}, it was first reported the existence of \textit{intertwined} discrete space-time translational symmetry(IDSTTS), characterized by combined temporal and spatial translations. Notably, this structure allows for scenarios where spatial translational symmetry may not exist at any given time, or temporal translational symmetry may be absent at any spatial location. A natural and profound question is whether this intertwined discrete spacetime symmetry can also be spontaneously broken, leading to the formation of a stable, ordered phase, i.e. the concept of “discrete spacetime crystals”(DSTC), in analogy to the phenomenon of pure discrete time crystals? 

\begin{figure}
	\centering
	\includegraphics[width=0.42\textwidth]{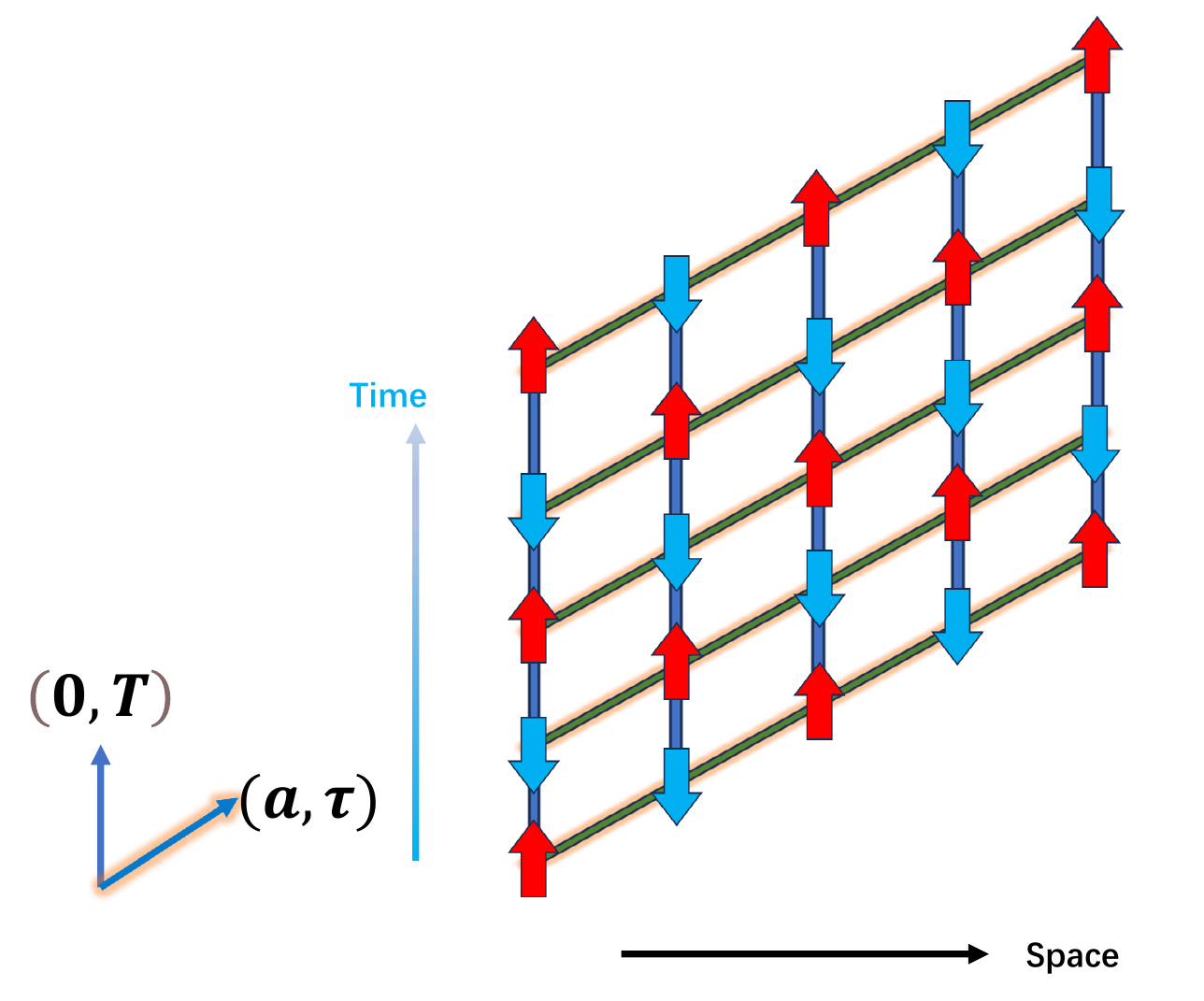}
	\caption{\textbf{Schematic diagram of intertwined discrete spacetime translational symmetry breaking.} Left shows the space-time primitive unit cell of the Hamiltonian constructed by the time periodicity $(0,T)$ and the intertwined spacetime periodicity $(a,\tau)$, which is space-time mixed. Right presents the spatial-temporal order of the spins. The period-doubling oscillations in both the directions indicate spontaneously breaking the IDSTTS.}\label{Fig.1}
\end{figure}

In this work, we attempt to answer this question by investigating a driven-dissipative classical spin chain. The presence of dissipation prevents the driving from heating the system towards a featureless infinite-temperature final state, and thus helps to stabilize the spacetime crystal order. Importantly, the classical nature of the spin system allows dissipation and perturbation (noise) to be easily added via a Landau-Lifshitz-Gilbert (LLG) equation, just as the classical Langevin dynamics. It is worth noting that the exotic non-equilibrium phases are not entirely dependent on the quantum or classical properties of the systems (e.g., the time crystal phases in classical many-body systems\cite{yao2020classical,yue2022thermal,lupo2019nanoscopic,gambetta2019classical,khasseh2019many,hurtado2020building,ye2021floquet,pizzi2021classical}, frustration induced spin ice physics in classical spin systems\cite{yue2023prethermal} etc. have been investigated). Here we demonstrate that the intertwined discrete spacetime symmetry breaking patterns can emerge in 1+1 spacetime dimensions due to the interplay between the novel driving and the dissipation, as shown in Fig.(\ref{Fig.1}).  Our finding stands in stark contrast to the previous studies on nonequilibrium spatial-temporal symmetry breaking phenomena focusing solely on direct product of spatial and temporal translational symmetries\cite{li2012space,smits2018observation,fan2024emergence,yue2023space,kleiner2021space,trager2021real}.

	\section { model}
 We consider a classical spin model in (1+1)-dimensional space-time that exhibits both the time periodicity $(0,T)$ and the intertwined spacetime periodicity $(a,\tau) $. The Hamiltonian reads
 
 	\begin{align}
   H=\sum_i J_i(t)s_i^xs_{i+1}^x+h_z\sum_is_i^z\label{1}
 	\end{align}
Which resembles a transverse Ising model but with driving interactions $J_i(t)$. Where the variable $\mathbf{s_i}=[s_i^x,s_i^y,s_i^z]$ is a three-dimensional classical vector with a fixed length $\lvert \mathbf{s_i}\rvert=\frac{1}{2}$. The driving interactions are given by $J_i(t)=J_0+\delta \mathrm{cos}[k(ia)-\omega_0t]$, where $J_0$ is set to be $1$ without losing generality, and $\delta$ is the tunable driving amplitude in units of $J_0$. We set $k=\frac{\omega_0 \tau}{a}$, then the system described has the following spatial-temporal translational symmetries: 
\begin{align}
	J_i(t+T)=J_i(t)=J_{i+1}(t+\tau)
\end{align}
 In the following, we set the time periodicity $T=\frac{2\pi}{\omega_0}=1$, and the lattice constant $a=1$ for simplicity.
Note that if $\tau$ is an irrational number, then there does not exist any pure spatial translational symmetry at any given time $t$. To simplify the analysis, we set $\tau\in[-0.5,0.5]$ so that $(1,\tau)$ is the minimum spacetime period. $h_z$ is the strength of a time independent transverse field, which is important for realizing nontrivial spin dynamics. Without this field, the spin dynamics would simply be a precession around the $x$ direction. In the absence of dissipation, the dynamics of each spin can be described by the equation of motion(EOM): $d\mathbf{s}_i/dt=\mathbf{s}_i\times\mathbf{H}_i$,where the effective magnetic field $\mathbf{H}_i=\nabla_\mathbf{s_i}H=[J_i(t)s_{i+1}^x+J_{i-1}(t)s_{i-1}^x,0,h_z]$. 

 The dissipation process in this system plays a crucial role in stabilizing the potential non-equilibrium phases of matter. It allows the system to relax towards a steady-state where the energy injected by the periodic driving is dissipated away, preventing the system from heating up and maintaining it in a non-equilibrium state. Furthermore, the fluctuation-dissipation theorem
 implies that dissipation must come with noise, which can be modeled by method similar to the classical Langevin dynamics. The EOM for each spin is described by a stochastic Landau-Lifshitz-Gilbert (LLG) equation\cite{lakshmanan2011fascinating,ma2010temperature}
\begin{align}
	\frac{d\mathbf{s}_i}{dt}=\mathbf{s}_i\times\widetilde{\mathbf{H}}_i+\lambda\mathbf{s}_i\times(\mathbf{s}_i\times\widetilde{\mathbf{H}}_i)\label{3}
\end{align}
where $\lambda$ is the dissipation strength fixed as $\lambda=1$ throughout this paper. And $\widetilde{\mathbf{H}}_i=\mathbf{H}_i+\boldsymbol{\xi}_i(t)$ is the effective magnetic field, where $\boldsymbol{\xi}_i(t)$ is a 3D zero-mean ($\langle \xi_i^{\alpha}(t)\rangle_{\boldsymbol{\xi}}=0$) random field representing thermal fluctuations. We further assume the local baths around different sites are independent of each other, and the stochastic variables satisfy $\langle \xi_i^{\alpha}(t)\xi_j^{\beta}(t^\prime)\rangle_{\boldsymbol{\xi}}=\mathcal{D}^2\delta_{ij}\delta_{\alpha\beta}\delta(t-t^\prime) $ where $\alpha,\beta=x,y,z$, $\mathcal{D}$ is the strength of the noise, and the average $\langle *\rangle_{\boldsymbol{\xi}}$ is over all the noise trajectories. If the bath is in thermal equilibrium with temperature $\mathrm{T}$ , the fluctuation dissipation theorem indicates that  $\mathcal{D}^2=2\lambda \mathrm{T}$. The stochastic differential Eq.(\ref{3}) is discretized by adopting Stratonovich’s formula, and solving it by the standard Heun method\cite{ament2016solving} with a time step of $\Delta t=10^{-3}$, the convergence of which has been checked numerically (see the Supplemental Material\cite{supplement}). The system size in our simulation is up to $L = 200$, which is more than enough for the simulation (see also the Supplemental Material\cite{supplement}).  
In the following, we will first focus on the long-time asymptotic dynamics of this model in the absence of noise perturbations and investigate the spontaneous symmetry breaking in spacetime.  Then we will show that the spatial-temporal order can persist up to an exponentially divergent lifetime in the presence of generic environmental noise.

	\section { Results and discussion}
	\subsection{Spatial-temporal order and intertwined discrete spacetime translational symmetry breaking}
	\begin{figure*}
		\centering
		\includegraphics[width=0.9\textwidth]{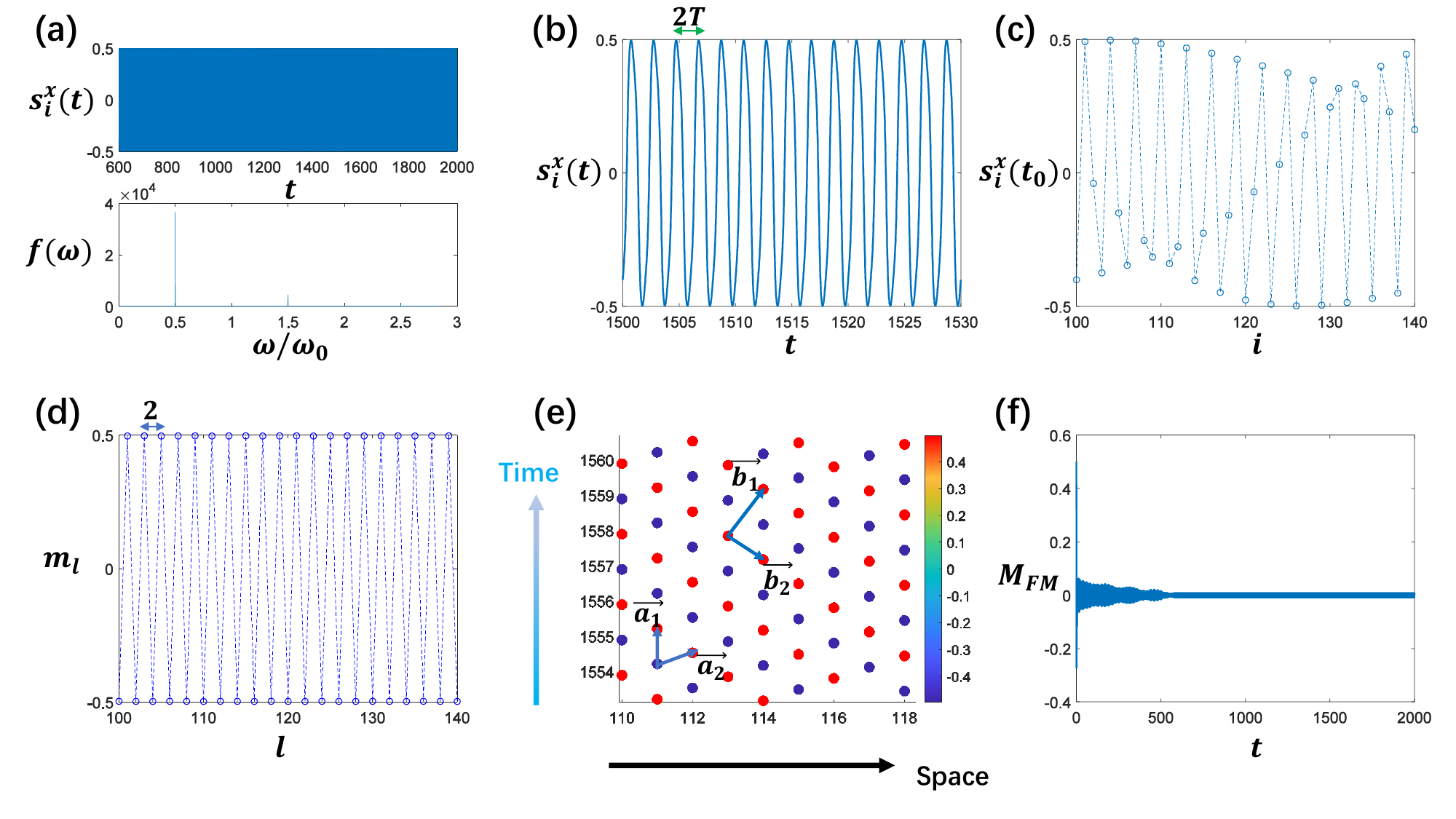}
		\caption{\textbf{Spatial-temporal order of the driven-dissipative spin system} (a) Long-time dynamics of $s_i^x$ on site $i$ (upper panel), and its Fourier spectrum (lower panel), $\omega_0=2\pi$.  (b) A detail of a smaller region in (a). (c) Snapshots of $s_i^x$ at time slice $t_0=1500$. (d) The plot of spacetime response, $m_l=s_l^x(t^\star+(l-1)\tau)$ with $t^\star=1500.2$. (e) Space-time symmetry breaking patterns of $s_i^x(t)$. $\vec a_1=(0,1)$ and $\vec a_2=(1,\tau)$ are primitive vectors of the Hamiltonian, while $\vec b_1=(1,\tau+1)$ and $\vec b_2=(1,\tau-1)$ are primitive vectors of the emergent DSTC. (f) Long-time dynamics of the ferromagnetic(FM) order parameter $M_{FM}(t)$. Parameters in our simulation are chosen as $L=200$, $\lambda=1$, $\mathcal{D}=0$, $h_z=2.5J_0$, $\delta=8J_0$,  $\tau=1/\pi$.  }\label{Fig.2}
	\end{figure*}
	\begin{figure*}
		\centering
		\includegraphics[width=0.9\textwidth]{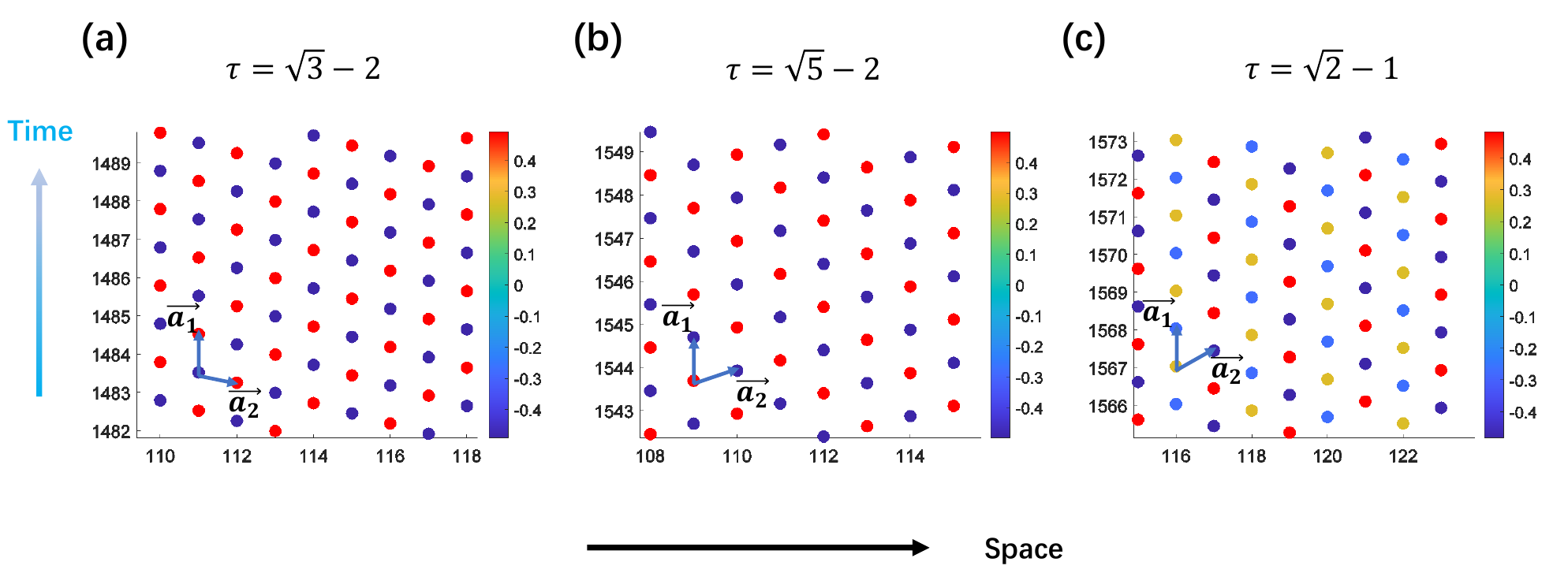}
		\caption{\textbf{Discrete spacetime crystals with other spacetime periods.} The parameters are chosen as  $\delta=8J_0$ for (a) and (b), and  $\delta=10J_0$ for (c). $L=200$, $\lambda=1$, $\mathcal{D}=0$, $h_z=2.5J_0$ for (a)-(c).   }\label{Fig.3}
	\end{figure*}
	We start the time evolution with a spatially uniform ferromagnetic initial state, i.e. $[\frac{1}{2},0,0]$ for all spins. However, we find that the long-time behavior does not depend on the initial state. (see Supplemental Material\cite{supplement} for comparison with an example starting from a nonuniform random initial state.)   We set $\tau=1/\pi$ so there is no independent spatial translational symmetry. In the following, we monitor the magnetization dynamics $s_i^x(t)$.   Fig.(\ref{Fig.2} a) is the long-time dynamics of $s_i^x(t)$ on site $i=100$ , and its Fourier spectrum reveals periodic doubling in time direction $(0,T)$. (It's worth noting that the phenomenon also occurs for other sites not located near the boundary). From Fig.(\ref{Fig.2} b), which provides a detailed plot of a smaller region in Fig.(\ref{Fig.2} a), we can observe that $s_i^x$ exhibits a persistent oscillation with a period of $2T$ ($T=1$). This periodic doubling phenomenon is a manifestation of spontaneous $Z_2$ discrete time
	translational symmetry breaking.
	
	 Fig.(\ref{Fig.2} d) depicts the magnetization response in spacetime direction  $(1,\tau)$ defined by $m_l=s_l^x(t_0^\prime+(l-1)\tau)$. It can be observed that $m_l$ returns to its original value after two periods of spacetime. We appropriately choose $t^\star=1500.2$ to ensure that $m_l$ reaches its maximum value ($\sim 0.5$), considering that $s^x_i(t)$ varies between $[-0.5,0.5]$. Fig.(\ref{Fig.2} e) is a plot of $s_i^x(t)$ in the two dimensional space-time, where the periodic doubling occurs not only in time, but also in the intertwined spacetime direction. It can be seen the emergent DSTC unit cell is also space-time mixed. On the other hand, the lack of pure spatial translational symmetry leads to a irregular distribution of $s_i^x$ at a fixed time (See Fig.(\ref{Fig.2} c)). Fig.(\ref{Fig.2} f) plots the ferromagnetic order parameter $M_{FM}=\frac{1}{L} \sum_i s_i^x(t)$ versus time. It can be seen that $M_{FM}$ fluctuates around zero ($<0.008$) after a sufficient long time, and it should decay to zero in the large system size limit $L\rightarrow \infty$.   
	
	Discrete spacetime crystals with other spacetime periods are presented in Fig.(\ref{Fig.3}) where $\tau$ is restricted within $[-0.5,0.5]$, such that $(1,\tau)$ corresponds to the minimum spacetime period. Typically, Fig.(\ref{Fig.3} a) and Fig.(\ref{Fig.3} b) recreate the SSB of IDSTTS observed in Fig.(\ref{Fig.2}), while Fig.(\ref{Fig.3} c) shows a more complex symmetry breaking pattern. Aside from a periodical doubling feature in both time $\vec a_1$ and spacetime $\vec a_2$ directions, there is a discrepancy in the amplitudes of $s_i^x$  between even and odd sites. This asymmetry may stem from the intricate interplay between the spacetime lattice structure, the coupling strengths, and the specific form of the dissipation present in the system.

	\begin{figure}
		\centering
		\includegraphics[width=0.5\textwidth]{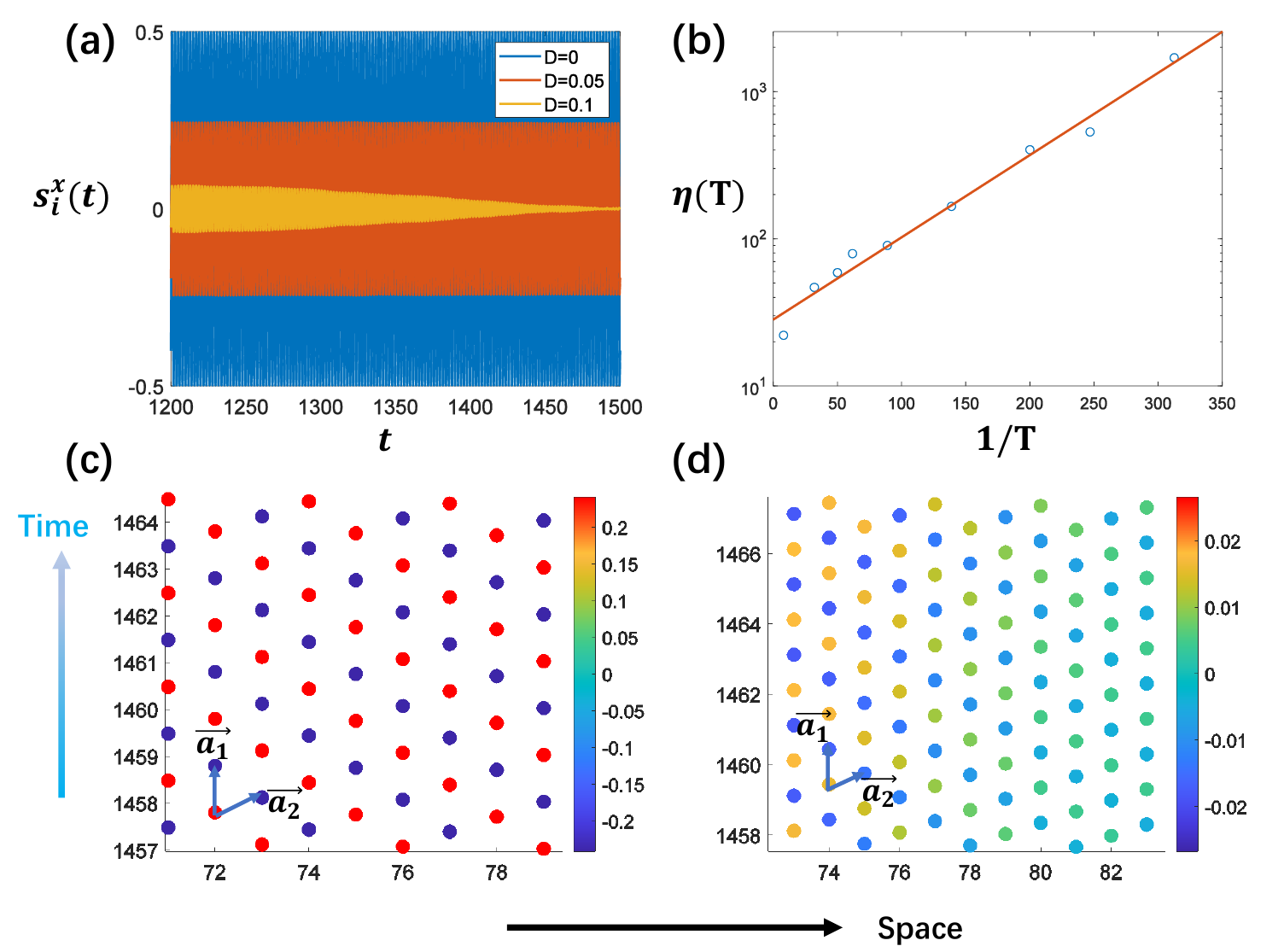}
		\caption{\textbf{Stability of the discrete spacetime crystal in presence of generic noises}  (a) Long-time dynamics of the average $\langle s_i^x(t)\rangle_{\boldsymbol{\xi}}$ on site $i$ after ensemble average over $2\times 10^3$ noise trajectories, with different noise strength. (b) The lifetime $\eta(\mathrm{T})$as a function of $\frac{1}{\mathrm{T}}$. (c) Activated Discrete spacetime crystal with noise strength of 0.05. (d) The melting of discrete spacetime crystal with noise strength of 0.1.   Other parameters are chosen the same as in Fig.(\ref{Fig.2}).   }\label{Fig.4}
	\end{figure}

	\subsection{Stability of the discrete spacetime crystal in presence of generic noises}
	We will analyze the stability of the proposed discrete spacetime crystal in presence of the noise perturbations of the dynamics. Environmental noises enter the EOM through Eq.(\ref{3}), perturbing the system continuously. In Fig.(\ref{Fig.4}a), we observe the long-time dynamics of $s_i^x(t)$ after ensemble averaging, where $\langle s_i^x(t)\rangle_{\boldsymbol{\xi}}$ exhibits a damped oscillation whose amplitude decays exponentially towards zero over time ($\sim e^{-\frac{t}{\eta}}$) when the noise strength $\mathcal{D}=0.1$. However, for $\mathcal{D}=0.05$, which corresponds to a relatively low temperature of the bath, thermal fluctuations are not strong enough to completely destroy the temporal order within the time scale of the simulation.  Fig.(\ref{Fig.4} d) depicts a "thermal melting" of the discrete spacetime crystal at   $\mathcal{D}=0.1$.  It should be note that due to the forbiddance of any finite-temperature phase transition for 1D spin chains, the proposed DSTC is actually an activated phase with lifetime that diverges exponentially as $\mathrm{T}\rightarrow0$, similar to the classical time crystal in Ref.\cite{yao2020classical}. The lifetime of the DSTC $\eta$ is a function of the temperature $\mathrm{T}$ ($\eta\sim e^{\frac{\Delta}{\mathrm{T}}}$), as shown in Fig.(\ref{Fig.4} b).  We can see that the spacetime order survives at a low temperature ($\mathcal{D}=0.05$) in the typical time scale of the simulation, which is accessible in cold atom experiments (See Fig.(\ref{Fig.4} c)).
	
	\subsection{Possible experimental realizations of our model } 
	Finally, we discuss the possible experimental realization of our model. A key challenge arises from the requirement for a periodic driving imposed on the spin-spin interactions rather than on the external field, which was also encountered by the previous works\cite{yue2022thermal,yue2023space,yue2023prethermal}. To address this challenge, two different experimental setups have been suggested by these works. One approach involves magnetophononics\cite{yarmohammadi2021dynamical,deltenre2021lattice,bossini2021femtosecond}, where the electric field of a laser is coupled to the phonon, leading to periodic atomic displacements that dynamically modulate the interactions between the spins. Another potential avenue for experimental realization is in synthetic quantum systems, such as trapped ions\cite{schneider2012experimental} and cavity quantum electrodynamics (QED) systems\cite{gopalakrishnan2009emergent}. Here, the magnetic interaction mediated by a cavity can be dynamically controlled by applying a periodic driving to the cavity photons, offering an alternative pathway for implementing the required driven interactions. It is important to note that the origin of SSB does not depend on the quantum or classical nature of the interacting systems, thus we expect the DSTC to be observed in the synthetic quantum systems. An open question is the effect of quantum fluctuations on the spacetime order. Therefore, a quantum generalization of our model would be a worthwhile direction to explore in future work. 
	 
	\section{Conclusion and Outlook}
	
	In conclusion, by investigating a driven-dissipative interacting spin system, we uncovered an exotic nonmagnetic phase of matter characterized by intertwined discrete spacetime translational symmetry breaking. This emergent phase, deeply rooted in the nonequilibrium dynamics of the system, establishes a distinctive long-range order not only in time but also in the intertwined spacetime direction. The unit cell of DSTC generally mixes space and time, which reminds us the equivalence between space and time in special relativity. However, in both quantum mechanics and classical dynamics they play different roles and even Lorentz invariance does not imply the complete equivalence between space and time.  Our results highlight the intimate relationship between these two dimensions from a perspective of SSB, where the system exhibits $Z_2$ symmetry breaking in the intertwined (mixed) spacetime direction.  
	
	Our work paves the way for several promising future developments. First, is it possible to realize DSTC in quantum systems, either closed or dissipative?  As previously discussed, the interplay between the quantum fluctuation and the spacetime order may give rise to rich dynamical behaviors.  Secondly, while our study preserved the Ising symmetry in Eq.(\ref{1}), it raises the question of whether similar exotic phases can be realized in systems with other symmetries, such as U(1) symmetry.  Another important direction involves studying the transition from DSTC to a phase without any spatial-temporal symmetry breaking. Reference\cite{yue2022thermal} used a similar method to achieve a DTC in a 3D classical driven-dissipative Ising model, where the phase transition has been systematically investigated. They found that the critical properties fall into the 3D Ising universality class, despite the genuine nonequilibrium nature of the driven system. Our model Eq.(\ref{1}) exists in 1+1D spacetime, and the  DSTC is an activated phase at finite temperature. This prompts the question of whether there may exist a DSTC with true long-range order and a critical finite temperature. Last but not least, can an intertwined space-time symmetry structure spontaneously emerge without any external driving?  This is a more fundamental and exciting question compared to breaking IDSTTS in our work and is certainly worth further investigation. 
	\section*{Acknowledgement}
	I thank Zi Cai for useful discussions. I also thank Shuohang Wu and Bo Fan for helpful comments on this work. This work is supported by the National Key Research and Development Program of China (Grant No. 2020YFA0309000), NSFC of  China (Grant No.12174251), Natural Science Foundation of Shanghai (Grant No.22ZR142830),  Shanghai Municipal Science and Technology Major Project (Grant No.2019SHZDZX01).
	\bibliography{spacetime.bib}

\begin{thebibliography}{42}
\expandafter\ifx\csname natexlab\endcsname\relax\def\natexlab#1{#1}\fi
\expandafter\ifx\csname bibnamefont\endcsname\relax
  \def\bibnamefont#1{#1}\fi
\expandafter\ifx\csname bibfnamefont\endcsname\relax
  \def\bibfnamefont#1{#1}\fi
\expandafter\ifx\csname citenamefont\endcsname\relax
  \def\citenamefont#1{#1}\fi
\expandafter\ifx\csname url\endcsname\relax
  \def\url#1{\texttt{#1}}\fi
\expandafter\ifx\csname urlprefix\endcsname\relax\def\urlprefix{URL }\fi
\providecommand{\bibinfo}[2]{#2}
\providecommand{\eprint}[2][]{\url{#2}}

\bibitem[{\citenamefont{Choi et~al.}(2017)\citenamefont{Choi, Choi, Landig,
  Kucsko, Zhou, Isoya, Jelezko, Onoda, Sumiya, Khemani
  et~al.}}]{choi2017observation}
\bibinfo{author}{\bibfnamefont{S.}~\bibnamefont{Choi}},
  \bibinfo{author}{\bibfnamefont{J.}~\bibnamefont{Choi}},
  \bibinfo{author}{\bibfnamefont{R.}~\bibnamefont{Landig}},
  \bibinfo{author}{\bibfnamefont{G.}~\bibnamefont{Kucsko}},
  \bibinfo{author}{\bibfnamefont{H.}~\bibnamefont{Zhou}},
  \bibinfo{author}{\bibfnamefont{J.}~\bibnamefont{Isoya}},
  \bibinfo{author}{\bibfnamefont{F.}~\bibnamefont{Jelezko}},
  \bibinfo{author}{\bibfnamefont{S.}~\bibnamefont{Onoda}},
  \bibinfo{author}{\bibfnamefont{H.}~\bibnamefont{Sumiya}},
  \bibinfo{author}{\bibfnamefont{V.}~\bibnamefont{Khemani}},
  \bibnamefont{et~al.}, \bibinfo{journal}{Nature}
  \textbf{\bibinfo{volume}{543}}, \bibinfo{pages}{221} (\bibinfo{year}{2017}).

\bibitem[{\citenamefont{Zhang et~al.}(2017)\citenamefont{Zhang, Hess,
  Kyprianidis, Becker, Lee, Smith, Pagano, Potirniche, Potter, Vishwanath
  et~al.}}]{zhang2017observation}
\bibinfo{author}{\bibfnamefont{J.}~\bibnamefont{Zhang}},
  \bibinfo{author}{\bibfnamefont{P.~W.} \bibnamefont{Hess}},
  \bibinfo{author}{\bibfnamefont{A.}~\bibnamefont{Kyprianidis}},
  \bibinfo{author}{\bibfnamefont{P.}~\bibnamefont{Becker}},
  \bibinfo{author}{\bibfnamefont{A.}~\bibnamefont{Lee}},
  \bibinfo{author}{\bibfnamefont{J.}~\bibnamefont{Smith}},
  \bibinfo{author}{\bibfnamefont{G.}~\bibnamefont{Pagano}},
  \bibinfo{author}{\bibfnamefont{I.-D.} \bibnamefont{Potirniche}},
  \bibinfo{author}{\bibfnamefont{A.~C.} \bibnamefont{Potter}},
  \bibinfo{author}{\bibfnamefont{A.}~\bibnamefont{Vishwanath}},
  \bibnamefont{et~al.}, \bibinfo{journal}{Nature}
  \textbf{\bibinfo{volume}{543}}, \bibinfo{pages}{217} (\bibinfo{year}{2017}).

\bibitem[{\citenamefont{Frey and Rachel}(2022)}]{frey2022realization}
\bibinfo{author}{\bibfnamefont{P.}~\bibnamefont{Frey}} \bibnamefont{and}
  \bibinfo{author}{\bibfnamefont{S.}~\bibnamefont{Rachel}},
  \bibinfo{journal}{Science advances} \textbf{\bibinfo{volume}{8}},
  \bibinfo{pages}{eabm7652} (\bibinfo{year}{2022}).

\bibitem[{\citenamefont{Autti et~al.}(2018)\citenamefont{Autti, Eltsov, and
  Volovik}}]{autti2018observation}
\bibinfo{author}{\bibfnamefont{S.}~\bibnamefont{Autti}},
  \bibinfo{author}{\bibfnamefont{V.}~\bibnamefont{Eltsov}}, \bibnamefont{and}
  \bibinfo{author}{\bibfnamefont{G.}~\bibnamefont{Volovik}},
  \bibinfo{journal}{Physical review letters} \textbf{\bibinfo{volume}{120}},
  \bibinfo{pages}{215301} (\bibinfo{year}{2018}).

\bibitem[{\citenamefont{Ke{\ss}ler et~al.}(2021)\citenamefont{Ke{\ss}ler,
  Kongkhambut, Georges, Mathey, Cosme, and Hemmerich}}]{kessler2021observation}
\bibinfo{author}{\bibfnamefont{H.}~\bibnamefont{Ke{\ss}ler}},
  \bibinfo{author}{\bibfnamefont{P.}~\bibnamefont{Kongkhambut}},
  \bibinfo{author}{\bibfnamefont{C.}~\bibnamefont{Georges}},
  \bibinfo{author}{\bibfnamefont{L.}~\bibnamefont{Mathey}},
  \bibinfo{author}{\bibfnamefont{J.~G.} \bibnamefont{Cosme}}, \bibnamefont{and}
  \bibinfo{author}{\bibfnamefont{A.}~\bibnamefont{Hemmerich}},
  \bibinfo{journal}{Physical Review Letters} \textbf{\bibinfo{volume}{127}},
  \bibinfo{pages}{043602} (\bibinfo{year}{2021}).

\bibitem[{\citenamefont{Mi et~al.}(2022)\citenamefont{Mi, Ippoliti, Quintana,
  Greene, Chen, Gross, Arute, Arya, Atalaya, Babbush et~al.}}]{mi2022time}
\bibinfo{author}{\bibfnamefont{X.}~\bibnamefont{Mi}},
  \bibinfo{author}{\bibfnamefont{M.}~\bibnamefont{Ippoliti}},
  \bibinfo{author}{\bibfnamefont{C.}~\bibnamefont{Quintana}},
  \bibinfo{author}{\bibfnamefont{A.}~\bibnamefont{Greene}},
  \bibinfo{author}{\bibfnamefont{Z.}~\bibnamefont{Chen}},
  \bibinfo{author}{\bibfnamefont{J.}~\bibnamefont{Gross}},
  \bibinfo{author}{\bibfnamefont{F.}~\bibnamefont{Arute}},
  \bibinfo{author}{\bibfnamefont{K.}~\bibnamefont{Arya}},
  \bibinfo{author}{\bibfnamefont{J.}~\bibnamefont{Atalaya}},
  \bibinfo{author}{\bibfnamefont{R.}~\bibnamefont{Babbush}},
  \bibnamefont{et~al.}, \bibinfo{journal}{Nature}
  \textbf{\bibinfo{volume}{601}}, \bibinfo{pages}{531} (\bibinfo{year}{2022}).

\bibitem[{\citenamefont{Tr{\"a}ger et~al.}(2021)\citenamefont{Tr{\"a}ger,
  Gruszecki, Lisiecki, Gro{\ss}, F{\"o}rster, Weigand, G{\l}owi{\'n}ski,
  Ku{\'s}wik, Dubowik, Sch{\"u}tz et~al.}}]{trager2021real}
\bibinfo{author}{\bibfnamefont{N.}~\bibnamefont{Tr{\"a}ger}},
  \bibinfo{author}{\bibfnamefont{P.}~\bibnamefont{Gruszecki}},
  \bibinfo{author}{\bibfnamefont{F.}~\bibnamefont{Lisiecki}},
  \bibinfo{author}{\bibfnamefont{F.}~\bibnamefont{Gro{\ss}}},
  \bibinfo{author}{\bibfnamefont{J.}~\bibnamefont{F{\"o}rster}},
  \bibinfo{author}{\bibfnamefont{M.}~\bibnamefont{Weigand}},
  \bibinfo{author}{\bibfnamefont{H.}~\bibnamefont{G{\l}owi{\'n}ski}},
  \bibinfo{author}{\bibfnamefont{P.}~\bibnamefont{Ku{\'s}wik}},
  \bibinfo{author}{\bibfnamefont{J.}~\bibnamefont{Dubowik}},
  \bibinfo{author}{\bibfnamefont{G.}~\bibnamefont{Sch{\"u}tz}},
  \bibnamefont{et~al.}, \bibinfo{journal}{Physical Review Letters}
  \textbf{\bibinfo{volume}{126}}, \bibinfo{pages}{057201}
  (\bibinfo{year}{2021}).

\bibitem[{\citenamefont{Zhu et~al.}(2019)\citenamefont{Zhu, Marino, Yao, Lukin,
  and Demler}}]{zhu2019dicke}
\bibinfo{author}{\bibfnamefont{B.}~\bibnamefont{Zhu}},
  \bibinfo{author}{\bibfnamefont{J.}~\bibnamefont{Marino}},
  \bibinfo{author}{\bibfnamefont{N.~Y.} \bibnamefont{Yao}},
  \bibinfo{author}{\bibfnamefont{M.~D.} \bibnamefont{Lukin}}, \bibnamefont{and}
  \bibinfo{author}{\bibfnamefont{E.~A.} \bibnamefont{Demler}},
  \bibinfo{journal}{New Journal of Physics} \textbf{\bibinfo{volume}{21}},
  \bibinfo{pages}{073028} (\bibinfo{year}{2019}).

\bibitem[{\citenamefont{Gong et~al.}(2018)\citenamefont{Gong, Hamazaki, and
  Ueda}}]{gong2018discrete}
\bibinfo{author}{\bibfnamefont{Z.}~\bibnamefont{Gong}},
  \bibinfo{author}{\bibfnamefont{R.}~\bibnamefont{Hamazaki}}, \bibnamefont{and}
  \bibinfo{author}{\bibfnamefont{M.}~\bibnamefont{Ueda}},
  \bibinfo{journal}{Physical review letters} \textbf{\bibinfo{volume}{120}},
  \bibinfo{pages}{040404} (\bibinfo{year}{2018}).

\bibitem[{\citenamefont{Yang and Cai}(2021)}]{yang2021dynamical}
\bibinfo{author}{\bibfnamefont{X.}~\bibnamefont{Yang}} \bibnamefont{and}
  \bibinfo{author}{\bibfnamefont{Z.}~\bibnamefont{Cai}},
  \bibinfo{journal}{Physical Review Letters} \textbf{\bibinfo{volume}{126}},
  \bibinfo{pages}{020602} (\bibinfo{year}{2021}).

\bibitem[{\citenamefont{Hu et~al.}(2023)\citenamefont{Hu, Fu, Li, and
  Shen}}]{hu2023solvable}
\bibinfo{author}{\bibfnamefont{Z.-A.} \bibnamefont{Hu}},
  \bibinfo{author}{\bibfnamefont{B.}~\bibnamefont{Fu}},
  \bibinfo{author}{\bibfnamefont{X.}~\bibnamefont{Li}}, \bibnamefont{and}
  \bibinfo{author}{\bibfnamefont{S.-Q.} \bibnamefont{Shen}},
  \bibinfo{journal}{Physical Review Research} \textbf{\bibinfo{volume}{5}},
  \bibinfo{pages}{L032024} (\bibinfo{year}{2023}).

\bibitem[{\citenamefont{Surace et~al.}(2019)\citenamefont{Surace, Russomanno,
  Dalmonte, Silva, Fazio, and Iemini}}]{surace2019floquet}
\bibinfo{author}{\bibfnamefont{F.~M.} \bibnamefont{Surace}},
  \bibinfo{author}{\bibfnamefont{A.}~\bibnamefont{Russomanno}},
  \bibinfo{author}{\bibfnamefont{M.}~\bibnamefont{Dalmonte}},
  \bibinfo{author}{\bibfnamefont{A.}~\bibnamefont{Silva}},
  \bibinfo{author}{\bibfnamefont{R.}~\bibnamefont{Fazio}}, \bibnamefont{and}
  \bibinfo{author}{\bibfnamefont{F.}~\bibnamefont{Iemini}},
  \bibinfo{journal}{Physical Review B} \textbf{\bibinfo{volume}{99}},
  \bibinfo{pages}{104303} (\bibinfo{year}{2019}).

\bibitem[{\citenamefont{Huang et~al.}(2018)\citenamefont{Huang, Wu, and
  Liu}}]{huang2018clean}
\bibinfo{author}{\bibfnamefont{B.}~\bibnamefont{Huang}},
  \bibinfo{author}{\bibfnamefont{Y.-H.} \bibnamefont{Wu}}, \bibnamefont{and}
  \bibinfo{author}{\bibfnamefont{W.~V.} \bibnamefont{Liu}},
  \bibinfo{journal}{Physical review letters} \textbf{\bibinfo{volume}{120}},
  \bibinfo{pages}{110603} (\bibinfo{year}{2018}).

\bibitem[{\citenamefont{Russomanno et~al.}(2017)\citenamefont{Russomanno,
  Iemini, Dalmonte, and Fazio}}]{russomanno2017floquet}
\bibinfo{author}{\bibfnamefont{A.}~\bibnamefont{Russomanno}},
  \bibinfo{author}{\bibfnamefont{F.}~\bibnamefont{Iemini}},
  \bibinfo{author}{\bibfnamefont{M.}~\bibnamefont{Dalmonte}}, \bibnamefont{and}
  \bibinfo{author}{\bibfnamefont{R.}~\bibnamefont{Fazio}},
  \bibinfo{journal}{Physical Review B} \textbf{\bibinfo{volume}{95}},
  \bibinfo{pages}{214307} (\bibinfo{year}{2017}).

\bibitem[{\citenamefont{Mu{\~n}oz-Arias
  et~al.}(2022)\citenamefont{Mu{\~n}oz-Arias, Chinni, and
  Poggi}}]{munoz2022floquet}
\bibinfo{author}{\bibfnamefont{M.~H.} \bibnamefont{Mu{\~n}oz-Arias}},
  \bibinfo{author}{\bibfnamefont{K.}~\bibnamefont{Chinni}}, \bibnamefont{and}
  \bibinfo{author}{\bibfnamefont{P.~M.} \bibnamefont{Poggi}},
  \bibinfo{journal}{Physical Review Research} \textbf{\bibinfo{volume}{4}},
  \bibinfo{pages}{023018} (\bibinfo{year}{2022}).

\bibitem[{\citenamefont{Yao et~al.}(2017)\citenamefont{Yao, Potter, Potirniche,
  and Vishwanath}}]{yao2017discrete}
\bibinfo{author}{\bibfnamefont{N.~Y.} \bibnamefont{Yao}},
  \bibinfo{author}{\bibfnamefont{A.~C.} \bibnamefont{Potter}},
  \bibinfo{author}{\bibfnamefont{I.-D.} \bibnamefont{Potirniche}},
  \bibnamefont{and}
  \bibinfo{author}{\bibfnamefont{A.}~\bibnamefont{Vishwanath}},
  \bibinfo{journal}{Physical review letters} \textbf{\bibinfo{volume}{118}},
  \bibinfo{pages}{030401} (\bibinfo{year}{2017}).

\bibitem[{\citenamefont{Collado et~al.}(2021)\citenamefont{Collado, Usaj,
  Balseiro, Zanette, and Lorenzana}}]{collado2021emergent}
\bibinfo{author}{\bibfnamefont{H.~O.} \bibnamefont{Collado}},
  \bibinfo{author}{\bibfnamefont{G.}~\bibnamefont{Usaj}},
  \bibinfo{author}{\bibfnamefont{C.~A.} \bibnamefont{Balseiro}},
  \bibinfo{author}{\bibfnamefont{D.~H.} \bibnamefont{Zanette}},
  \bibnamefont{and}
  \bibinfo{author}{\bibfnamefont{J.}~\bibnamefont{Lorenzana}},
  \bibinfo{journal}{Physical Review Research} \textbf{\bibinfo{volume}{3}},
  \bibinfo{pages}{L042023} (\bibinfo{year}{2021}).

\bibitem[{\citenamefont{von Keyserlingk et~al.}(2016)\citenamefont{von
  Keyserlingk, Khemani, and Sondhi}}]{von2016absolute}
\bibinfo{author}{\bibfnamefont{C.~W.} \bibnamefont{von Keyserlingk}},
  \bibinfo{author}{\bibfnamefont{V.}~\bibnamefont{Khemani}}, \bibnamefont{and}
  \bibinfo{author}{\bibfnamefont{S.~L.} \bibnamefont{Sondhi}},
  \bibinfo{journal}{Physical Review B} \textbf{\bibinfo{volume}{94}},
  \bibinfo{pages}{085112} (\bibinfo{year}{2016}).

\bibitem[{\citenamefont{Xu and Wu}(2018)}]{xu2018space}
\bibinfo{author}{\bibfnamefont{S.}~\bibnamefont{Xu}} \bibnamefont{and}
  \bibinfo{author}{\bibfnamefont{C.}~\bibnamefont{Wu}},
  \bibinfo{journal}{Physical Review Letters} \textbf{\bibinfo{volume}{120}},
  \bibinfo{pages}{096401} (\bibinfo{year}{2018}).

\bibitem[{\citenamefont{Yao et~al.}(2020)\citenamefont{Yao, Nayak, Balents, and
  Zaletel}}]{yao2020classical}
\bibinfo{author}{\bibfnamefont{N.~Y.} \bibnamefont{Yao}},
  \bibinfo{author}{\bibfnamefont{C.}~\bibnamefont{Nayak}},
  \bibinfo{author}{\bibfnamefont{L.}~\bibnamefont{Balents}}, \bibnamefont{and}
  \bibinfo{author}{\bibfnamefont{M.~P.} \bibnamefont{Zaletel}},
  \bibinfo{journal}{Nature Physics} \textbf{\bibinfo{volume}{16}},
  \bibinfo{pages}{438} (\bibinfo{year}{2020}).

\bibitem[{\citenamefont{Yue et~al.}(2022)\citenamefont{Yue, Yang, and
  Cai}}]{yue2022thermal}
\bibinfo{author}{\bibfnamefont{M.}~\bibnamefont{Yue}},
  \bibinfo{author}{\bibfnamefont{X.}~\bibnamefont{Yang}}, \bibnamefont{and}
  \bibinfo{author}{\bibfnamefont{Z.}~\bibnamefont{Cai}},
  \bibinfo{journal}{Physical Review B} \textbf{\bibinfo{volume}{105}},
  \bibinfo{pages}{L100303} (\bibinfo{year}{2022}).

\bibitem[{\citenamefont{Lupo and Weber}(2019)}]{lupo2019nanoscopic}
\bibinfo{author}{\bibfnamefont{C.}~\bibnamefont{Lupo}} \bibnamefont{and}
  \bibinfo{author}{\bibfnamefont{C.}~\bibnamefont{Weber}},
  \bibinfo{journal}{Physical Review B} \textbf{\bibinfo{volume}{100}},
  \bibinfo{pages}{195431} (\bibinfo{year}{2019}).

\bibitem[{\citenamefont{Gambetta et~al.}(2019)\citenamefont{Gambetta, Carollo,
  Lazarides, Lesanovsky, and Garrahan}}]{gambetta2019classical}
\bibinfo{author}{\bibfnamefont{F.}~\bibnamefont{Gambetta}},
  \bibinfo{author}{\bibfnamefont{F.}~\bibnamefont{Carollo}},
  \bibinfo{author}{\bibfnamefont{A.}~\bibnamefont{Lazarides}},
  \bibinfo{author}{\bibfnamefont{I.}~\bibnamefont{Lesanovsky}},
  \bibnamefont{and} \bibinfo{author}{\bibfnamefont{J.}~\bibnamefont{Garrahan}},
  \bibinfo{journal}{Physical Review E} \textbf{\bibinfo{volume}{100}},
  \bibinfo{pages}{060105} (\bibinfo{year}{2019}).

\bibitem[{\citenamefont{Khasseh et~al.}(2019)\citenamefont{Khasseh, Fazio,
  Ruffo, and Russomanno}}]{khasseh2019many}
\bibinfo{author}{\bibfnamefont{R.}~\bibnamefont{Khasseh}},
  \bibinfo{author}{\bibfnamefont{R.}~\bibnamefont{Fazio}},
  \bibinfo{author}{\bibfnamefont{S.}~\bibnamefont{Ruffo}}, \bibnamefont{and}
  \bibinfo{author}{\bibfnamefont{A.}~\bibnamefont{Russomanno}},
  \bibinfo{journal}{Physical review letters} \textbf{\bibinfo{volume}{123}},
  \bibinfo{pages}{184301} (\bibinfo{year}{2019}).

\bibitem[{\citenamefont{Hurtado-Guti{\'e}rrez
  et~al.}(2020)\citenamefont{Hurtado-Guti{\'e}rrez, Carollo,
  P{\'e}rez-Espigares, and Hurtado}}]{hurtado2020building}
\bibinfo{author}{\bibfnamefont{R.}~\bibnamefont{Hurtado-Guti{\'e}rrez}},
  \bibinfo{author}{\bibfnamefont{F.}~\bibnamefont{Carollo}},
  \bibinfo{author}{\bibfnamefont{C.}~\bibnamefont{P{\'e}rez-Espigares}},
  \bibnamefont{and} \bibinfo{author}{\bibfnamefont{P.}~\bibnamefont{Hurtado}},
  \bibinfo{journal}{Physical Review Letters} \textbf{\bibinfo{volume}{125}},
  \bibinfo{pages}{160601} (\bibinfo{year}{2020}).

\bibitem[{\citenamefont{Ye et~al.}(2021)\citenamefont{Ye, Machado, and
  Yao}}]{ye2021floquet}
\bibinfo{author}{\bibfnamefont{B.}~\bibnamefont{Ye}},
  \bibinfo{author}{\bibfnamefont{F.}~\bibnamefont{Machado}}, \bibnamefont{and}
  \bibinfo{author}{\bibfnamefont{N.~Y.} \bibnamefont{Yao}},
  \bibinfo{journal}{Physical Review Letters} \textbf{\bibinfo{volume}{127}},
  \bibinfo{pages}{140603} (\bibinfo{year}{2021}).

\bibitem[{\citenamefont{Pizzi et~al.}(2021)\citenamefont{Pizzi, Nunnenkamp, and
  Knolle}}]{pizzi2021classical}
\bibinfo{author}{\bibfnamefont{A.}~\bibnamefont{Pizzi}},
  \bibinfo{author}{\bibfnamefont{A.}~\bibnamefont{Nunnenkamp}},
  \bibnamefont{and} \bibinfo{author}{\bibfnamefont{J.}~\bibnamefont{Knolle}},
  \bibinfo{journal}{Physical Review Letters} \textbf{\bibinfo{volume}{127}},
  \bibinfo{pages}{140602} (\bibinfo{year}{2021}).

\bibitem[{\citenamefont{Yue and Cai}(2023{\natexlab{a}})}]{yue2023prethermal}
\bibinfo{author}{\bibfnamefont{M.}~\bibnamefont{Yue}} \bibnamefont{and}
  \bibinfo{author}{\bibfnamefont{Z.}~\bibnamefont{Cai}},
  \bibinfo{journal}{Physical Review Letters} \textbf{\bibinfo{volume}{131}},
  \bibinfo{pages}{056502} (\bibinfo{year}{2023}{\natexlab{a}}).

\bibitem[{\citenamefont{Li et~al.}(2012)\citenamefont{Li, Gong, Yin, Quan, Yin,
  Zhang, Duan, and Zhang}}]{li2012space}
\bibinfo{author}{\bibfnamefont{T.}~\bibnamefont{Li}},
  \bibinfo{author}{\bibfnamefont{Z.-X.} \bibnamefont{Gong}},
  \bibinfo{author}{\bibfnamefont{Z.-Q.} \bibnamefont{Yin}},
  \bibinfo{author}{\bibfnamefont{H.}~\bibnamefont{Quan}},
  \bibinfo{author}{\bibfnamefont{X.}~\bibnamefont{Yin}},
  \bibinfo{author}{\bibfnamefont{P.}~\bibnamefont{Zhang}},
  \bibinfo{author}{\bibfnamefont{L.-M.} \bibnamefont{Duan}}, \bibnamefont{and}
  \bibinfo{author}{\bibfnamefont{X.}~\bibnamefont{Zhang}},
  \bibinfo{journal}{Physical review letters} \textbf{\bibinfo{volume}{109}},
  \bibinfo{pages}{163001} (\bibinfo{year}{2012}).

\bibitem[{\citenamefont{Smits et~al.}(2018)\citenamefont{Smits, Liao, Stoof,
  and van~der Straten}}]{smits2018observation}
\bibinfo{author}{\bibfnamefont{J.}~\bibnamefont{Smits}},
  \bibinfo{author}{\bibfnamefont{L.}~\bibnamefont{Liao}},
  \bibinfo{author}{\bibfnamefont{H.}~\bibnamefont{Stoof}}, \bibnamefont{and}
  \bibinfo{author}{\bibfnamefont{P.}~\bibnamefont{van~der Straten}},
  \bibinfo{journal}{Physical review letters} \textbf{\bibinfo{volume}{121}},
  \bibinfo{pages}{185301} (\bibinfo{year}{2018}).

\bibitem[{\citenamefont{Fan et~al.}(2024)\citenamefont{Fan, Cai, and
  Garc{\'\i}a-Garc{\'\i}a}}]{fan2024emergence}
\bibinfo{author}{\bibfnamefont{B.}~\bibnamefont{Fan}},
  \bibinfo{author}{\bibfnamefont{Z.}~\bibnamefont{Cai}}, \bibnamefont{and}
  \bibinfo{author}{\bibfnamefont{A.~M.} \bibnamefont{Garc{\'\i}a-Garc{\'\i}a}},
  \bibinfo{journal}{arXiv preprint arXiv:2405.14216}  (\bibinfo{year}{2024}).

\bibitem[{\citenamefont{Yue and Cai}(2023{\natexlab{b}})}]{yue2023space}
\bibinfo{author}{\bibfnamefont{M.}~\bibnamefont{Yue}} \bibnamefont{and}
  \bibinfo{author}{\bibfnamefont{Z.}~\bibnamefont{Cai}},
  \bibinfo{journal}{Physical Review B} \textbf{\bibinfo{volume}{107}},
  \bibinfo{pages}{094313} (\bibinfo{year}{2023}{\natexlab{b}}).

\bibitem[{\citenamefont{Kleiner et~al.}(2021)\citenamefont{Kleiner, Zhou,
  Dorsch, Zhang, Koelle, and Jin}}]{kleiner2021space}
\bibinfo{author}{\bibfnamefont{R.}~\bibnamefont{Kleiner}},
  \bibinfo{author}{\bibfnamefont{X.}~\bibnamefont{Zhou}},
  \bibinfo{author}{\bibfnamefont{E.}~\bibnamefont{Dorsch}},
  \bibinfo{author}{\bibfnamefont{X.}~\bibnamefont{Zhang}},
  \bibinfo{author}{\bibfnamefont{D.}~\bibnamefont{Koelle}}, \bibnamefont{and}
  \bibinfo{author}{\bibfnamefont{D.}~\bibnamefont{Jin}},
  \bibinfo{journal}{Nature Communications} \textbf{\bibinfo{volume}{12}},
  \bibinfo{pages}{6038} (\bibinfo{year}{2021}).

\bibitem[{\citenamefont{Lakshmanan}(2011)}]{lakshmanan2011fascinating}
\bibinfo{author}{\bibfnamefont{M.}~\bibnamefont{Lakshmanan}},
  \bibinfo{journal}{Philosophical Transactions of the Royal Society A:
  Mathematical, Physical and Engineering Sciences}
  \textbf{\bibinfo{volume}{369}}, \bibinfo{pages}{1280} (\bibinfo{year}{2011}).

\bibitem[{\citenamefont{Ma et~al.}(2010)\citenamefont{Ma, Dudarev, Semenov, and
  Woo}}]{ma2010temperature}
\bibinfo{author}{\bibfnamefont{P.-W.} \bibnamefont{Ma}},
  \bibinfo{author}{\bibfnamefont{S.}~\bibnamefont{Dudarev}},
  \bibinfo{author}{\bibfnamefont{A.}~\bibnamefont{Semenov}}, \bibnamefont{and}
  \bibinfo{author}{\bibfnamefont{C.}~\bibnamefont{Woo}},
  \bibinfo{journal}{Physical Review E} \textbf{\bibinfo{volume}{82}},
  \bibinfo{pages}{031111} (\bibinfo{year}{2010}).

\bibitem[{\citenamefont{Ament et~al.}(2016)\citenamefont{Ament, Rangarajan,
  Parthasarathy, and Rakheja}}]{ament2016solving}
\bibinfo{author}{\bibfnamefont{S.}~\bibnamefont{Ament}},
  \bibinfo{author}{\bibfnamefont{N.}~\bibnamefont{Rangarajan}},
  \bibinfo{author}{\bibfnamefont{A.}~\bibnamefont{Parthasarathy}},
  \bibnamefont{and} \bibinfo{author}{\bibfnamefont{S.}~\bibnamefont{Rakheja}},
  \bibinfo{journal}{arXiv preprint arXiv:1607.04596}  (\bibinfo{year}{2016}).

\bibitem[{\citenamefont{see supplementary material at [URL]~for
  details.}()}]{supplement}
\bibinfo{author}{\bibnamefont{see supplementary material at [URL]~for
  details.}}

\bibitem[{\citenamefont{Yarmohammadi et~al.}(2021)\citenamefont{Yarmohammadi,
  Meyer, Fauseweh, Normand, and Uhrig}}]{yarmohammadi2021dynamical}
\bibinfo{author}{\bibfnamefont{M.}~\bibnamefont{Yarmohammadi}},
  \bibinfo{author}{\bibfnamefont{C.}~\bibnamefont{Meyer}},
  \bibinfo{author}{\bibfnamefont{B.}~\bibnamefont{Fauseweh}},
  \bibinfo{author}{\bibfnamefont{B.}~\bibnamefont{Normand}}, \bibnamefont{and}
  \bibinfo{author}{\bibfnamefont{G.}~\bibnamefont{Uhrig}},
  \bibinfo{journal}{Physical Review B} \textbf{\bibinfo{volume}{103}},
  \bibinfo{pages}{045132} (\bibinfo{year}{2021}).

\bibitem[{\citenamefont{Deltenre et~al.}(2021)\citenamefont{Deltenre, Bossini,
  Anders, and Uhrig}}]{deltenre2021lattice}
\bibinfo{author}{\bibfnamefont{K.}~\bibnamefont{Deltenre}},
  \bibinfo{author}{\bibfnamefont{D.}~\bibnamefont{Bossini}},
  \bibinfo{author}{\bibfnamefont{F.~B.} \bibnamefont{Anders}},
  \bibnamefont{and} \bibinfo{author}{\bibfnamefont{G.~S.} \bibnamefont{Uhrig}},
  \bibinfo{journal}{Physical Review B} \textbf{\bibinfo{volume}{104}},
  \bibinfo{pages}{184419} (\bibinfo{year}{2021}).

\bibitem[{\citenamefont{Bossini et~al.}(2021)\citenamefont{Bossini, Dal~Conte,
  Terschanski, Springholz, Bonanni, Deltenre, Anders, Uhrig, Cerullo, and
  Cinchetti}}]{bossini2021femtosecond}
\bibinfo{author}{\bibfnamefont{D.}~\bibnamefont{Bossini}},
  \bibinfo{author}{\bibfnamefont{S.}~\bibnamefont{Dal~Conte}},
  \bibinfo{author}{\bibfnamefont{M.}~\bibnamefont{Terschanski}},
  \bibinfo{author}{\bibfnamefont{G.}~\bibnamefont{Springholz}},
  \bibinfo{author}{\bibfnamefont{A.}~\bibnamefont{Bonanni}},
  \bibinfo{author}{\bibfnamefont{K.}~\bibnamefont{Deltenre}},
  \bibinfo{author}{\bibfnamefont{F.}~\bibnamefont{Anders}},
  \bibinfo{author}{\bibfnamefont{G.~S.} \bibnamefont{Uhrig}},
  \bibinfo{author}{\bibfnamefont{G.}~\bibnamefont{Cerullo}}, \bibnamefont{and}
  \bibinfo{author}{\bibfnamefont{M.}~\bibnamefont{Cinchetti}},
  \bibinfo{journal}{Physical Review B} \textbf{\bibinfo{volume}{104}},
  \bibinfo{pages}{224424} (\bibinfo{year}{2021}).

\bibitem[{\citenamefont{Schneider et~al.}(2012)\citenamefont{Schneider, Porras,
  and Schaetz}}]{schneider2012experimental}
\bibinfo{author}{\bibfnamefont{C.}~\bibnamefont{Schneider}},
  \bibinfo{author}{\bibfnamefont{D.}~\bibnamefont{Porras}}, \bibnamefont{and}
  \bibinfo{author}{\bibfnamefont{T.}~\bibnamefont{Schaetz}},
  \bibinfo{journal}{Reports on Progress in Physics}
  \textbf{\bibinfo{volume}{75}}, \bibinfo{pages}{024401}
  (\bibinfo{year}{2012}).

\bibitem[{\citenamefont{Gopalakrishnan
  et~al.}(2009)\citenamefont{Gopalakrishnan, Lev, and
  Goldbart}}]{gopalakrishnan2009emergent}
\bibinfo{author}{\bibfnamefont{S.}~\bibnamefont{Gopalakrishnan}},
  \bibinfo{author}{\bibfnamefont{B.~L.} \bibnamefont{Lev}}, \bibnamefont{and}
  \bibinfo{author}{\bibfnamefont{P.~M.} \bibnamefont{Goldbart}},
  \bibinfo{journal}{Nature Physics} \textbf{\bibinfo{volume}{5}},
  \bibinfo{pages}{845} (\bibinfo{year}{2009}).

\end{thebibliography}

	\end{document}